\documentstyle[prl,aps,twocolumn,amssymb,psfig]{revtex}

\begin{document}
\draft


\noindent
{\bf Comment on ``Quantum Suppression of Shot Noise in 
Atom-Size Metallic Contacts''}

\vspace{4mm}

In a recent letter \cite{brom}, van den Brom and van Ruitenbeek
found a pronounced suppression of the shot noise 
in atom-size gold contacts
with conductances near integer multiples of 
$G_0=2e^2/h$, revealing unambiguously the quantized nature of the electronic
transport. 
However, the ad hoc model they introduced to describe the contribution
of partially-open conductance channels to the shot noise is unable to fit 
either the maxima or minima of their shot noise data.
%
Here we point out that a 
model of quantum-confined electrons with disorder \cite{burki} 
quantitatively reproduces the measurements of Ref.\ \onlinecite{brom}. 

We model a 
nanocontact in a monovalent metal as 
a deformable constriction in an 
electron gas,
with disorder included via randomly distributed delta-function potentials
\cite{burki}. 
For convenience, the system is taken to be two-dimensional.
The transmission probabilities $T_n$ of the conducting channels
are obtained via a modified recursive Green's function algorithm
\cite{burki}. 
The dimensionless shot noise $s_I$ at zero temperature
is \cite{buttiker}
\begin{equation}
s_I \equiv \frac{P_I}{2eI}=\frac{\sum_n T_n(1-T_n)}{\sum_n T_n},
\label{noise}
\end{equation}
where $P_I$ is the shot noise spectral density and
$I$ is the time-average current. 
Plotting $s_I$ versus the conductance $G=G_0\sum_n T_n$ 
eliminates the dependence on dimensionality
for an ideal contact, provided no special symmetries are present.

\begin{figure}
\begin{center}
\psfig{figure=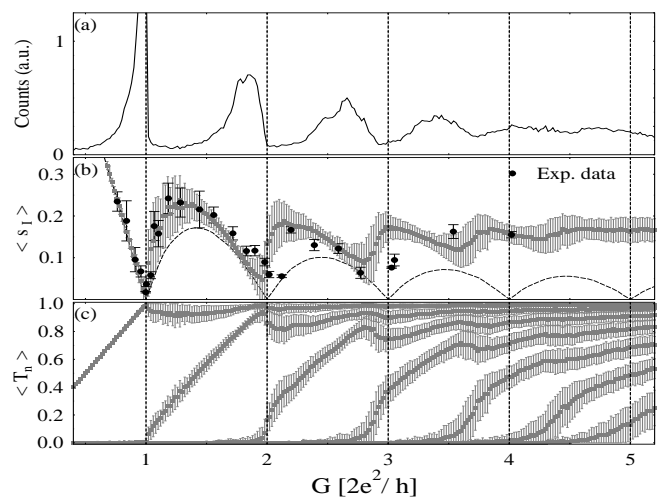,width=8.6cm,height=6.5cm,angle=-90}
\end{center}
\vspace{-0.5cm}
\caption{
(a) Conductance histogram reproduced from  
Ref.\ [2]; (b)
calculated mean shot noise $\left<s_I\right>$ (grey squares),
together with experimental data from Ref.\ [1] (black circles); 
(c) mean transmission probabilities $\left< T_n\right>$. 
The error bars indicate the 
standard deviations of the numerical results over the ensemble
and the experimental errors, respectively.}
\label{fig}
\end{figure}

Starting from the numerical
data that were used to generate the conductance histogram 
in Ref.\ \cite{burki}, we compute the mean and standard deviation
of $s_I$ and $T_n$ as functions of $G$.
The averages are taken over an ensemble of impurity configurations and contact
shapes.  
The agreement of the experimental results for particular contacts
and the calculated distribution of $s_I$ shown in 
Fig.\ \ref{fig}(b) 
is extremely good: 67\% of the 
experimental points lie within one standard deviation of 
$\langle s_I \rangle$ and 89\% lie within two standard deviations \cite{temp}.
It should be emphasized that no attempt has been made to fit the shot-noise
data; the numerical data of Ref.\ \onlinecite{burki}, where the 
length of the contact and the
strength of the disorder (mean-free path $k_F \ell=70$) were chosen to give 
qualitative agreement with experimental conductance histograms for
gold \cite{hist}, 
have simply been re-analyzed to calculate $\langle s_I \rangle$.

Contrary to the model of Ref.\ \cite{brom}, 
the minima of $\langle s_I\rangle$ do not occur at
integer multiples of $G_0$, but are shifted to lower values, which 
correspond {\em not} to maxima of the conductance histogram, but
rather to maxima of $\langle T_n \rangle$.

Fig.\ \ref{fig}(c) shows that the number of partially-open channels 
increases in proportion to $G$.
For comparison, the shot noise for a contact with only one
partially-open channel, which sets a lower bound,
is shown as a dashed curve in Fig.\ \ref{fig}(b).
The presence of several partially open channels for $G> G_0$
increases $\langle s_I \rangle$ above this lower bound, 
leading to an apparent saturation
at $\langle s_I \rangle \approx 0.18$ for larger contacts.
Neither the maxima nor the minima of the 
experimental data can be fit 
by the model of Ref.\ \onlinecite{brom}, which includes
only two partially-open conductance channels.

The excellent agreement between the shot-noise data of Ref.\ \onlinecite{brom}
and our model calculation suggests that quantum transport in gold nanocontacts
can be well described by a model
which includes only two essential features, quantum confinement and coherent
backscattering from imperfections in the contact.

J.\ B.\ acknowledges support from Swiss National Foundation 
PNR 36 ``Nanosciences'' grant \# 4036-044033. 

\vspace{4mm}

\noindent
J. B\"urki$^{1,2,3}$ and C.~A. Stafford$^1$

$^1$University of Arizona, Tucson, Arizona 85721

$^2$Universit\'{e} de Fribourg, 1700 Fribourg, Switzerland

$^3$IRRMA, EPFL, 1015 Lausanne, Switzerland

\vspace{4mm}

\noindent
Submitted to Phys. Rev. Lett. on 3 June 1999


\begin{references}
\bibitem{temp}
A finite temperature slightly lowers $s_I$ for 
larger $G$, but does not significantly change the quality of the fit.
\end{references}
\end{document}